\documentclass[doublecol]{epl2}
\usepackage[T1]{fontenc}
\usepackage[latin9]{inputenc}
\usepackage{amsmath}
\usepackage{esint}
\makeatletter
\@ifundefined{textcolor}{}
{%
 \definecolor{BLACK}{gray}{0}
 \definecolor{WHITE}{gray}{1}
 \definecolor{RED}{rgb}{1,0,0}
 \definecolor{GREEN}{rgb}{0,1,0}
 \definecolor{BLUE}{rgb}{0,0,1}
 \definecolor{CYAN}{cmyk}{1,0,0,0}
 \definecolor{MAGENTA}{cmyk}{0,1,0,0}
 \definecolor{YELLOW}{cmyk}{0,0,1,0}
 }

\title{(Sr$_3$La$_2$O$_{5}$)(Zn$_{1-x}$Mn$_x$)$_2$As$_2$: A Bulk Form Diluted Magnetic
Semiconductor isostructural to the "32522" Fe-based Superconductors}
\shorttitle{A Bulk Form Diluted Magnetic Semiconductor
(Sr$_3$La$_2$O$_{5}$)(Zn$_{1-x}$Mn$_x$)$_2$As$_2$}

\author{Huiyuan Man\inst{1}, Chuan Qin\inst{1}, Cui Ding\inst{1}, Quan Wang\inst{1}, Xin Gong\inst{1}, Shengli Guo\inst{1}, Hangdong Wang\inst{2}, Bin Chen\inst{2}, and F.L. Ning\inst{1}\footnote{Electronic address: ningfl@zju.edu.cn}}
\shortauthor{Huiyuan Man \etal}

 \institute{
  \inst{1} Department of Physics, Zhejiang University - Hangzhou 310027, China\\
  \inst{2} Department of Physics, Hangzhou Normal University - Hangzhou 310016, China
}


\abstract {A new diluted magnetic semiconductor system,
(Sr$_3$La$_2$O$_{5}$)(Zn$_{1-x}$Mn$_x$)$_2$As$_2$, has been
synthesized and characterized. 10\% Mn substitution for Zn in bulk
form (Sr$_3$La$_2$O$_{5}$)Zn$_2$As$_2$ results in a ferromagnetic
ordering below Curie temperature, $T_C$ $\sim$ 40 K.
(Sr$_3$La$_2$O$_{5}$)(Zn$_{1-x}$Mn$_x$)$_2$As$_2$ has a layered
crystal structure identical to that of 32522-type Fe-based
superconductors, and represents the fifth DMS family that has a
direct counterpart among the FeAs high temperature superconductor
families.}

\begin{document}

\maketitle

\section{Introduction}
The research of DMS (diluted magnetic semiconductors) has been
explosive following the successful fabrication of III-V (Ga,Mn)As
ferromagnetic thin films by Ohno et al in 1990's\cite{Ohno}. Over
the past two decades, much progress has been made in the fabrication
of DMS materials and the understanding of the ferromagnetism
\cite{Samarth,Chambers,Dietl1,Zutic,Jungwirth}. On the other hand,
most extensively studied DMS materials are thin films that are grown
under non-equilibrium condition, which encounters some inherent
difficulties. For example in (Ga,Mn)As, some Mn impurities enter the
interstitial sites, and makes it difficult to precisely determine
the amount of Mn that substitutes ionic Ga, which donates a hole and
acts as a local moment \cite{Jungwirth}. The thin films also
prohibit the utilization of powerful magnetic probes such as neutron
scattering and nuclear magnetic resonance (NMR) that are based on
bulk form specimens, to provide complementary information for
understanding the ferromagnetism at a microscopic level. Seeking for
bulk form DMS system grown in thermally equilibrium condition will
be helpful to understand the ferromagnetism.

Recently, through doping Mn into the I-II-V semiconductors LiZnAs
and LiZnP, Deng et al successfully synthesized two bulk DMS systems,
Li(Zn,Mn)As \cite{Deng1} and Li(Zn,Mn)P \cite{Deng2}, with $T_C$
$\sim$ 50 K. The I-II-V DMSs have advantages of decoupling spins and
carriers, where spins are introduced by Mn atoms and carriers are
created by off-stoichiometry of Li concentrations. This advantage
makes it possible to precisely control the amount of spins and
carriers, and investigate their individual effects on the
ferromagnetic ordering. More recently, several more bulk DMS systems
have been reported. Firstly, Ding et al reported the ferromagnetic
ordering below $T_C$ $\sim$ 40 K in a ``1111" type (La,Ba)(Zn,Mn)AsO
system \cite{Ding1}; Han et al reported the ferromagnetism in
(La,Ca)(Zn,Mn)SbO semiconductor \cite{HanW} and Yang et al reported
the fabrication of (La,Sr)(Cu,Mn)SO DMS with $T_C$ $\sim$ 210 K
\cite{Yangxj}. Secondly, Zhao et al. reported the ``122" type DMS
systems, (Ba,K)(Zn,Mn)$_{2}$As$_{2}$, which has $T_C$ as high as 180
K \cite{ZhaoK}, and Yang et al observed the ferromagnetic transition
below $T_C$ $\sim$ 17 K and a large negative magnetoresistance in
(Ba,K)(Cd,Mn)$_{2}$As$_{2}$ \cite{Yang2}. The Curie temperature of
(La,Sr)(Cu,Mn)SO and (Ba,K)(Zn,Mn)$_{2}$As$_{2}$ polycrystals is
already comparable to the record $T_C$ of (Ga,Mn)As thin films
\cite{ZhaoJH1, ZhaoJH2}.

The availability of bulk form DMS specimens readily enables the
microscopic investigation by $\mu$SR (muon spin relaxation) and NMR
techniques. $\mu$SR has demonstrated that the exchange interaction
supporting ferromagnetic coupling in Li(Zn,Mn)As, (La,Ba)(Zn,Mn)AsO,
(Ba,K)(Zn,Mn)$_{2}$As$_{2}$ and (Ga,Mn)As has a common origin and
comparable magnitude for a given spatial density of ordered moments,
no matter the specimens are thin films or bulk forms \cite{Deng1,
ZhaoK, Ding1, Dunsiger}. Moreover, Ding et al has conducted $^7$Li
NMR measurement of Li(Zn,Mn)P, and successfully detected the fast
relaxed Li sites that have Mn at nearest neighbor sites. They found
that the Mn spin-spin interactions extend over many unit cells,
which explains why DMSs could exhibit a relatively high $T_C$ with
such a low density of Mn \cite{Ding2}.

More interestingly, each of above ``111", ``1111" and ``122" DMSs
families has a direct counterpart among the FeAs-based high
temperature superconductor families. For example, ``1111"-type
(La,Ba)(Zn,Mn)AsO DMS has a ZrCuSiAs-type tetragonal structure,
identical to FeAs-based ``1111" type LaFeAsO$_{1-x}$F$_x$ high
temperature superconductor ($T_c$ = 26 K) \cite{Kamihara} and the
antiferromagnetic LaMnAsO ($T_N$ = 317 K) \cite{Emery}. The
excellent lattice matching (lattice constants are within 5\%
difference) between ferromagnetic, antiferrmagnetic and
superconducting systems opens the possibilities to make junctions
between these systems through the As layer. The parameters for all
these compounds are listed in Table. 1. We also mention that for the
``11" type Fe-based superconductor FeSe \cite{WuMK}, the counterpart
is (Zn,Mn)Se, which has been extensively studied as one of the
prototypical II-VI DMSs.

In this letter, we report successful synthesis and characterization
of a new bulk DMS system,
(Sr$_3$La$_2$O$_{5}$)(Zn$_{1-x}$Mn$_x$)$_2$As$_2$, which is
iso-structural to the ``32522" FeAs-based superconductor
(Ca$_3$Al$_2$O$_{5}$)Fe$_2$As$_2$ \cite{Shirage}. 5\% Mn
substitution for Zn in the parent semiconductor
(Sr$_3$La$_2$O$_{5}$)Zn$_2$As$_2$ results in ferromagnetic ordering,
as indicated by the strong enhancement of magnetization below $T_C$
$\sim$ 36 K. The bifurcation of ZFC and FC curves below $T_f$ = 12.5
K and a parallelogram-shaped hysteresis loop are also observed.
$T_C$ increases to 40 K with the doping level increasing to 10\%,
but starts to decrease at the doping level of 20\%. The saturation
moment is suppressed from 0.5 $\mu_B$/Mn for $x$ = 0.10 to 0.2
$\mu_B$/Mn for $x$ = 0.20. More Mn doping also suppresses the
coercive field from 324 mT for $x$ = 0.05 to 111 mT for $x$ = 0.20.
No ferromagnetic ordering is observed for the doping level of 30\%
Mn.

\begin{table*}[t]
  \begin{center}
    \begin{tabular}{|c|c|c|c|} \hline
      \emph -- &{SC} &{DMS}&{AFM}  \\
        \hline
      \emph&{FeSe}&{(Zn,Mn)Se} &{MnSe}\\
        &($T_c$$\sim$8K)\cite{WuMK} & ($T_f$$\sim$24K)\cite{Twardowski}
        &($T_N$$\sim$197K)\cite{Ito}\\
        {"11"}&Tetragonal&Cubic &Cubic\\
        &a=3.7676{\AA}&a=5.669{\AA}&a=5.464{\AA}\\
        &c=5.4847{\AA}&&\\
        \hline
      \emph&{LiFeAs}&{Li(Zn,Mn)As}&{LiMnAs}\\
        &($T_c$$\sim$18K)\cite{Wangxc}&($T_C$$\sim$50K)\cite{Deng1}
        &($T_N$$\sim$393K)\cite{Bronger}\\
        {"111"}&Tetragonal&Cubic&Tetragonal\\
        &a=3.77{\AA}&a=5.94{\AA}&a=4.273{\AA}\\
        &c=6.36{\AA}& &c=12.370{\AA}\\
        \hline
      \emph&{LaFeAs(O,F)}&{(La,Ba)(Zn,Mn)AsO}&{LaMnAsO}\\
        &($T_c$$\sim$26K)\cite{Kamihara}&($T_C$$\sim$40K)\cite{Ding1}
        &($T_N$$\sim$317K)\cite{Emery}\\
        {"1111"}&Tetragonal&Tetragonal&Tetragonal\\
        &a=4.0320{\AA}&a=4.116{\AA}&a=4.11398{\AA}\\
        &c=8.7263{\AA}&c=9.11{\AA}& c=9.03044{\AA}\\
        \hline
      \emph&{(Ba,K)Fe$_2$As$_2$} &{(Ba,K)(Zn,Mn)$_2$As$_2$}&{BaMn$_2$As$_2$}\\
        &($T_c$$\sim$ 38K)\cite{Rotter}&($T_C$$\sim$ 180K)\cite{ZhaoK}
        &($T_N$$\sim$ 625K)\cite{Singh}\\
        {"122"}&Tetragonal&Tetragonal&Tetragonal\\
        &a=3.917{\AA}&a=4.131{\AA}&a=4.1684{\AA}\\
        &c=13.2968{\AA}&c=13.481{\AA}& c=13.4681{\AA}\\
        \hline
      \emph&{Ca$_3$Al$_2$O$_{5-y}$Fe$_2$As$_2$}&{Sr$_3$La$_2$O$_5$(Zn,Mn)$_2$As$_2$}&\\
        &($T_c$$\sim$ 30.2K)\cite{Shirage}&($T_C$$\sim$ 40K,this work)&{hypothetical}\\
        {"32522"}&Tetragonal&Tetragonal&(Sr$_3$La$_2$O$_5$Mn$_2$As$_2$)\\
        &a=3.742{\AA}&a=4.2612{\AA}& \\
        &c=26.078{\AA}&c=27.675{\AA} & \\
        \hline
      \emph&{Sr$_4$V$_2$O$_6$Fe$_2$As$_2$}&{Sr$_4$Ti$_2$O$_6$(Zn,Mn)$_2$As$_2$}&\\
         & (T$_c$$\sim$ 37.2K)\cite{Zhu}&$(T_C$$\sim$ 25K)&{hypothetical}\\
         {"42622"}&Tetragonal &(Unpublished)&(Sr$_4$Ti$_2$O$_6$Mn$_2$As$_2$)\\
         &a=3.9296{\AA}& & \\
         &c=15.6732{\AA}& & \\
         \hline
    \end{tabular}
    \caption{The transition temperature $T_c$ for a superconductor (SC), Curie temperature $T_C$ for a diluted magnetic semiconductor (DMS) and Neel temperature $T_N$ for an antiferromagnet (AFM) of ``11", ``111",
    ``1111", ``122", ``32522" and ``42622" type compounds. The type of crystal structure and lattice constants are also listed for available compounds.}
  \end{center}
\end{table*}

\section{Experimental methods}

We synthesized (Sr$_3$La$_2$O$_{5}$)(Zn$_{1-x}$Mn$_x$)$_2$As$_2$
($x$ = 0.00, 0.05, 0.10, 0.20, 0.30) polycrystalline specimens by
the solid state reaction method. High purity elements of La, Zn, Mn
and As were mixed and heated to 900 $^{\circ}$C in an evacuated
silica tube to produce intermediate products LaAs, ZnAs and MnAs.
They were then mixed with La$_2$O$_3$, SrO with nominal
concentrations and slowly heated up to 1150 $^{\circ}$C, and held
for 50 hours before cooling to room temperature with turning off the
furnace. The polycrystals were characterized by X-ray diffraction at
room temperature and dc magnetization by Quantum Design SQUID. The
electrical resistance was measured on sintered pellets with typical
four-probe method.

\section{Results and discussion}

We show the crystal structure of
(Sr$_3$La$_2$O$_{5}$)(Zn$_{1-x}$Mn$_x$)$_2$As$_2$ and the X-ray
diffraction patterns in Fig. 1. Appreciable Bragg peaks from the
parent compound (Sr$_3$La$_2$O$_{5}$)(Zn$_2$As$_2$) can be indexed
by a layered tetragonal crystal structure ($I_4/mmm$), with a =
4.2612 {\AA} and c = 27.675 {\AA}. These lattice constants are close
to a = 4.069 {\AA} and  c = 26.876 {\AA} of a ``32522" compound
(Sr$_3$Sc$_2$O$_{5}$)(Fe$_2$As$_2$)\cite{WenHH}, as well as a =
3.742 {\AA} and c = 26.078 {\AA} of the superconducting sample
(Ca$_3$Al$_2$O$_{5}$)Fe$_2$As$_2$\cite{Shirage}. We observed
secondary phases of Zn$_3$As$_2$ and Sr$_2$As for $x$ $\leq$ 0.10,
as marked by the stars and the arrows in Fig.1. These impurities are
non-magnetic, which will not affect the magnetic properties
discussed in the following section.

\begin{figure}[!htpb] \centering \vspace*{+0.5cm}
\centering
\includegraphics[width=3.3in]{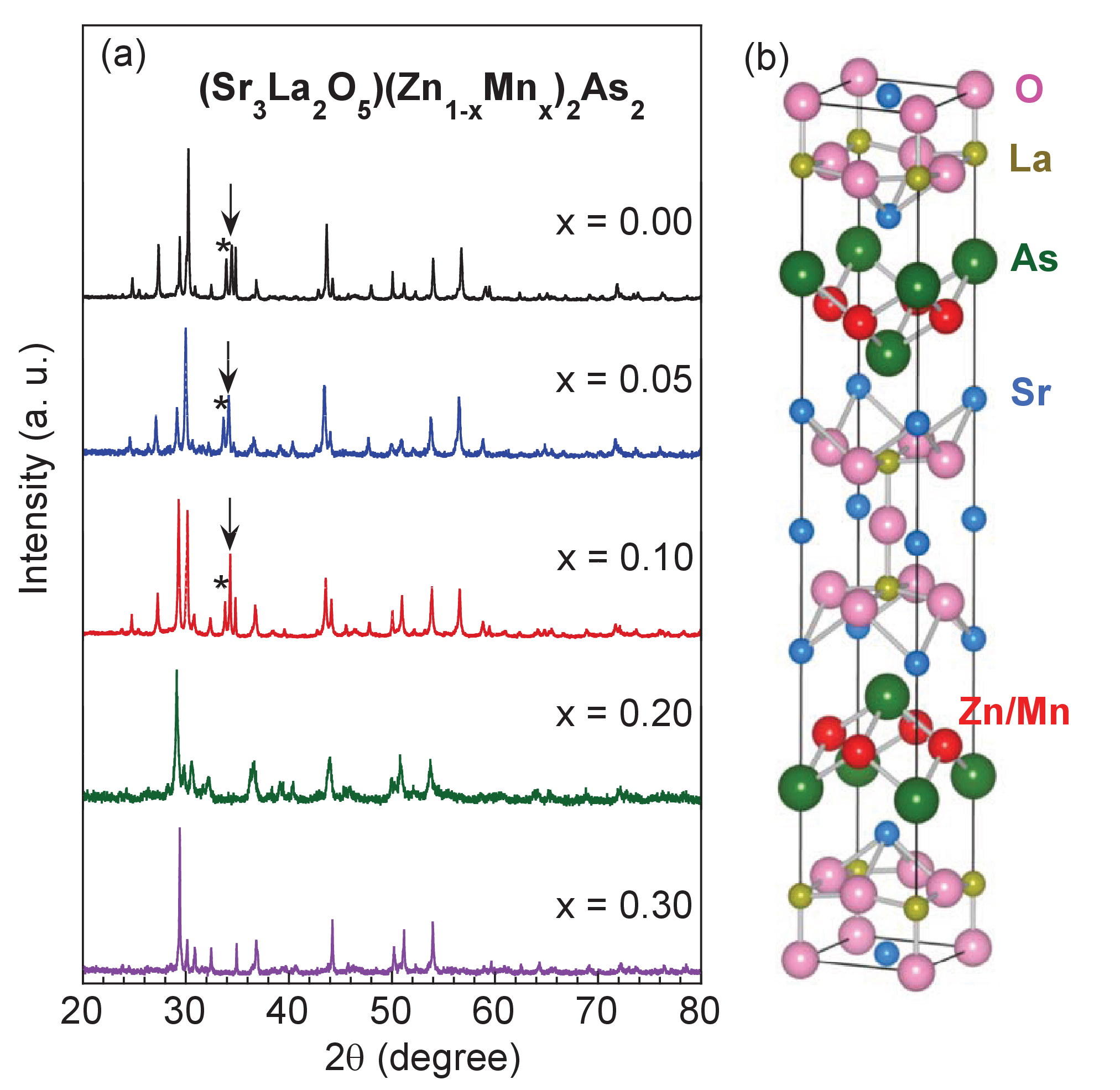}\vspace*{+0.5cm}
\caption{\label{Fig1:epsart} (Color online) X-ray diffraction
pattern (a) and crystal structure (b) of
(Sr$_3$La$_2$O$_{5}$)(Zn$_{1-x}$Mn$_x$)$_2$As$_2$. Secondary phases
of Sr$_2$As ($\downarrow$) and Zn$_3$As$_2$ ($\ast$) are marked for
$x$ $\leq$ 0.10.}
\end{figure}

In Fig. 2, we show the electrical resistivity measured for
(Sr$_3$La$_2$O$_{5}$)(Zn$_{1-x}$Mn$_x$)$_2$As$_2$ with $x$ = 0.05,
0.10, 0.20, 0.30.  The resistivity of all samples monotonically
increases toward lower temperature. This type of behavior in Mn
doped specimens has been observed in heavily doped region of
(Ga$_{1-x}$Mn$_{x}$)As \cite{Jungwirth}, as well as
(La,Ba)(Zn,Mn)AsO \cite{Ding1} and (Ba,K)(Zn,Mn)$_{2}$As$_{2}$
\cite{ZhaoK} DMS polycrystals. It has been ascribed to the
scattering of carriers by the magnetic fluctuations through exchange
interactions in (Ga$_{1-x}$Mn$_{x}$)As \cite{Jungwirth}. We have
also conducted the Hall effect measurements for the sample of
(Sr$_3$La$_2$O$_{5}$)(Zn$_{0.90}$Mn$_{0.10}$)$_2$As$_2$. The large
resistivity forbids us to accurately determine the carriers type,
and a preliminary measurement indicates that the carrier density is
in the order of $10^{16}$ cm$^{-3}$. This carrier density is
comparable to that of Li$_{1.1}$Zn$_{1-x}$Mn$_{x}$P \cite{Deng2} but
4 orders smaller than that of Li$_{1.1}$Zn$_{1-x}$Mn$_{x}$As
\cite{Deng1}.

\begin{figure}[!htpb] \centering \vspace*{+0.2cm}
\centering
\includegraphics[width=3.3in]{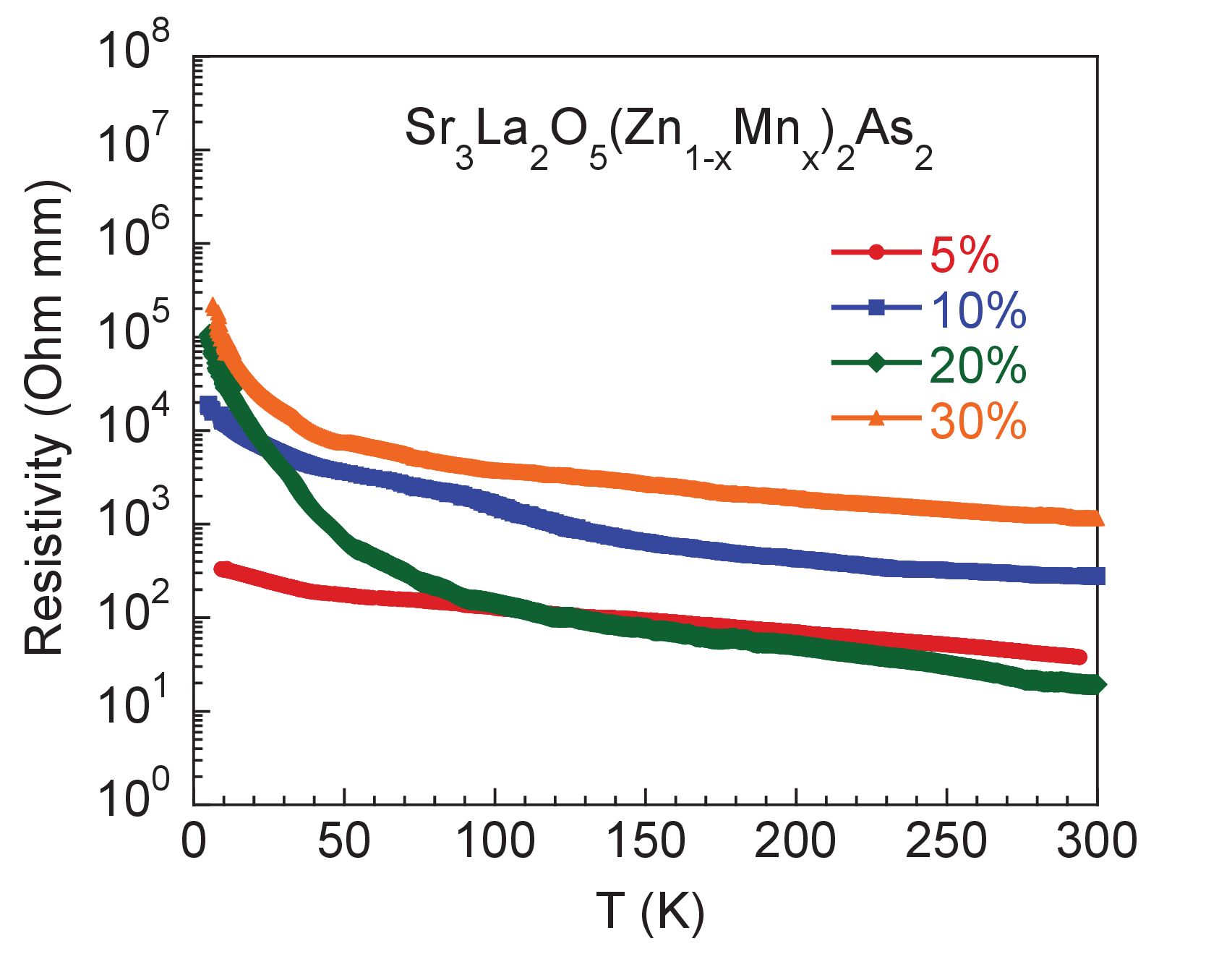}\vspace*{+0.1cm}
\caption{\label{Fig2:epsart} (Color online) The electrical
resistivity of (Sr$_3$La$_2$O$_{5}$)(Zn$_{1-x}$Mn$_x$)$_2$As$_2$
with $x$ = 0.05, 0.10, 0.20, 0.30.}
\end{figure}

In Fig. 3, we show the zero-field cooled (ZFC) and field cooled (FC)
measurements of the $dc$-magnetization $M$ of
(Sr$_3$La$_2$O$_{5}$)(Zn$_{1-x}$Mn$_x$)$_2$As$_2$ for $B_{ext}$ =
0.1 Tesla. For the doping of $x$ = 0.05, we observe a strong
increase of M at $T_C$ = 36 K, and the bifurcation of ZFC and FC
curves below the temperature $T_f$ = 12.5 K, where $T_f$ stands for
the freezing temperature of individual spins or domain wall motion.
The saturation moment at 2 K is 0.4 $\mu_B$/Mn. With the doping
level increasing to $x$ = 0.10, $T_C$ increases to $\sim$ 40 K, and
the saturation moment increases to 0.5 $\mu_B$/Mn. This indicates
that additional Mn atoms raise the ferromagnetic ordering
temperature. Both $T_C$ and moment size start to decrease with
further doping to $x$ = 0.20. The ferromagnetic ordering disappears
for the doping level of $x$ = 0.30. This is primarily due to the
competition of antiferromagnetic exchange interaction between spins
from nearest neighbor Mn sites. For 100\% Mn substitution for Zn in
``11", ``111", ``1111" and ``122" compounds, the ending product is
always an antiferromagnet with Neel temperature $T_N$ $\sim$ 200 K -
600 K, as displayed in Table. 1. Following this trend, we would
expect a hypothetical antiferromagnet with ``32522" structure,
(Sr$_3$La$_2$O$_{5}$)Mn$_2$As$_2$. We fit the temperature dependence
of $M$ above $T_C$ to a Curie-Weiss law. The effective paramagnetic
moment is determined to be 5 $\sim$ 6$\mu _B$/Mn, indicating that
the valence of Mn ions are +2, and it is in a high spin state, as
observed in other Mn doped DMSs \cite{Jungwirth, Zutic, Deng1}.

\begin{figure}[!htpb] \centering \vspace*{-0.1cm}
\centering
\includegraphics[width=3.3in]{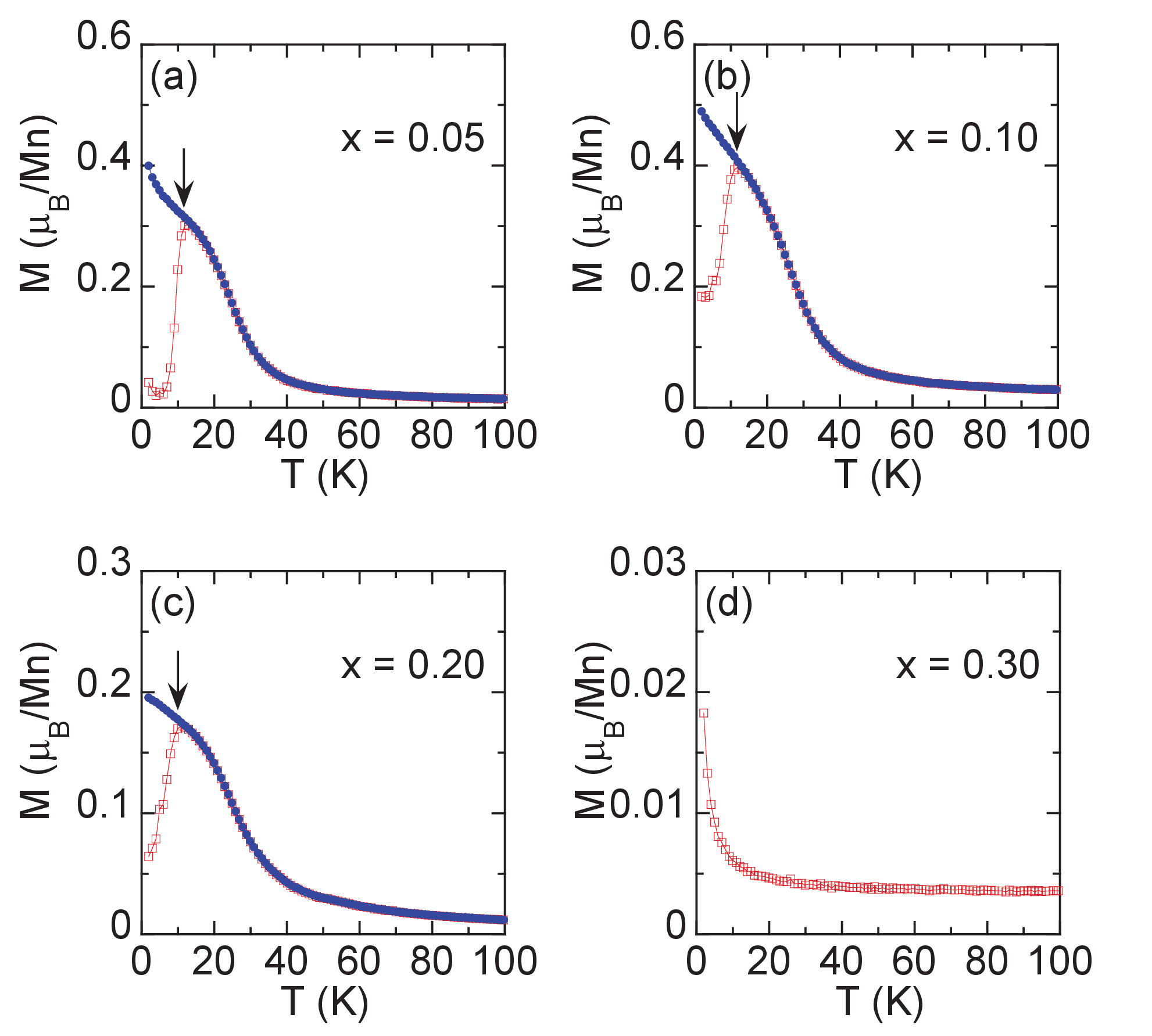}\vspace*{-0.1cm} \\
\caption{\label{Fig3:epsart} (Color online)  The magnetization $M$
for (Sr$_3$La$_2$O$_{5}$)(Zn$_{1-x}$Mn$_x$)$_2$As$_2$ with $x$ =
0.05, 0.10, 0.20, 0.30 obtained in the zero field cooling (ZFC, in
red) and field cooling (FC, in blue) mode under the external field
of 0.1 Tesla. Arrows mark the position of $T_f$. We note that $T_C$
is defined as the temperature where M shows a sharp upturn in the
low field of 1 mT (not shown), higher magnetic field suppresses the
sharp upturn feature.}
\end{figure}

In Fig. 4, we show the isothermal magnetization of
(Sr$_3$La$_2$O$_{5}$)(Zn$_{1-x}$Mn$_x$)$_2$As$_2$. For $x$ = 0.05, a
parallelogram-shaped hysteresis loop with a coercive field of 324 mT
is observed at 5 K. The coercive field continuously decreases to 111
mT with the doping level increasing to $x$ = 0.20. The coercive
fields are larger than $\sim$ 5 - 10 \revision{mT} of the cubic
structural (Ga$_{0.965}$Mn$_{0.035}$)As \cite{Ohno},
Li$_{1.1}$(Zn$_{0.97}$Mn$_{0.03}$)As \cite{Deng1}, and
Li$_{1.1}$(Zn$_{0.97}$Mn$_{0.03}$)P \cite{Deng2}, but smaller than
$\sim$1000 mT of the two dimensional (La,Ba)(Zn,Mn)AsO \cite{Ding1}
and (Ba,K)(Zn,Mn)$_{2}$As$_{2}$\cite{ZhaoK}. We show the temperature
dependence of the hysteresis loop for $x$ = 0.10 in Fig. 4(d). The
coercive field decreases from 203 mT to 42 mT at 10 K and becomes
zero at 60 K.

\begin{figure}[!htpb] \centering \vspace*{+0.1cm}
\centering
\includegraphics[width=3in]{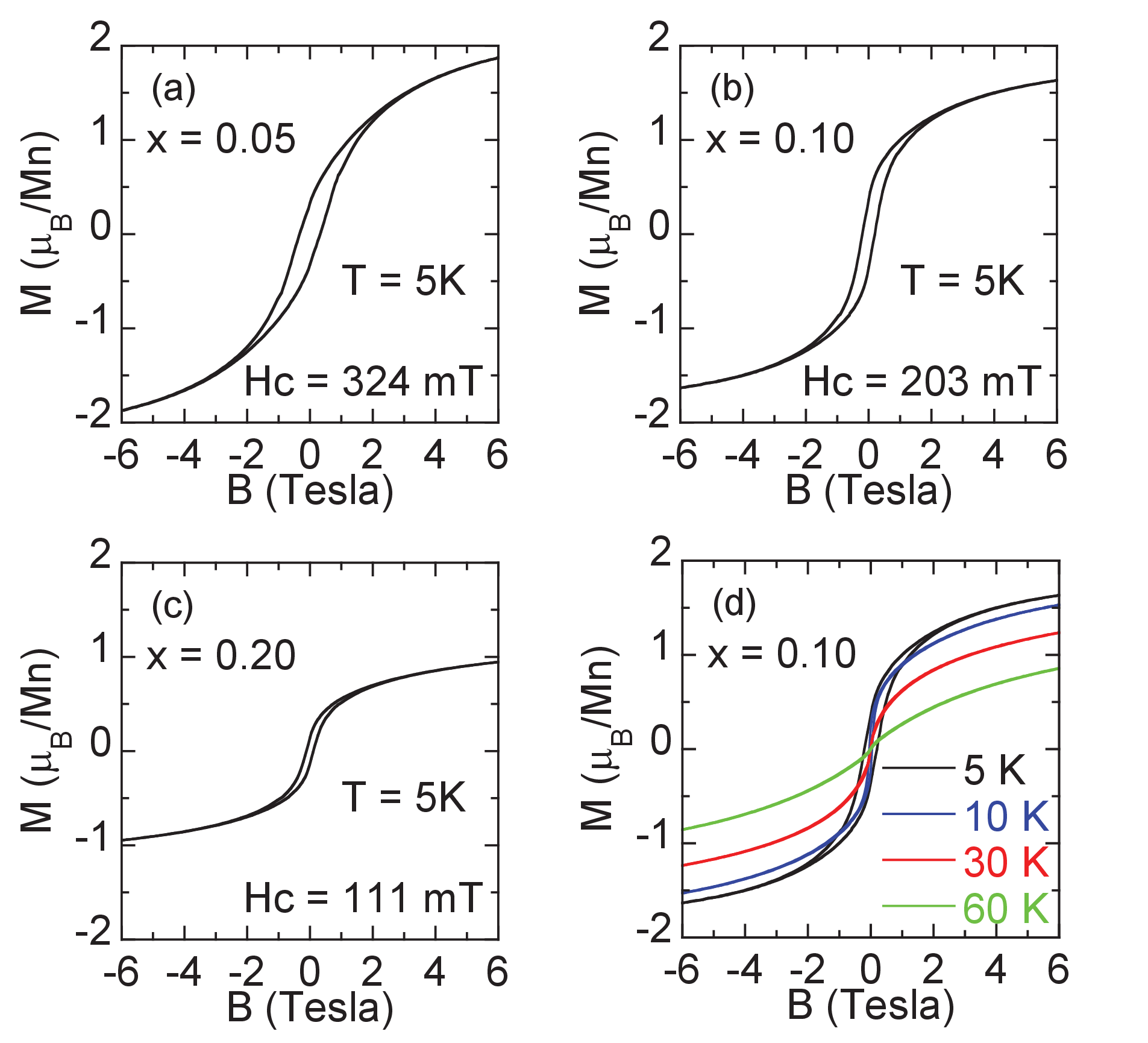}\vspace*{+0.1cm}
\caption{\label{Fig4:epsart} (Color online) The isothermal
magnetization for (Sr$_3$La$_2$O$_{5}$)(Zn$_{1-x}$Mn$_x$)$_2$As$_2$
with $x$ = 0.05, 0.10, 0.20 measured at 5 K, and for $x$ = 0.10
measured at 5 K, 10 K, 30 K and 60 K. $H_c$ values are the coercive
filed at 5 K.}
\end{figure}

As we have discussed in ref. \cite{Ding1}, the bifurcation of ZFC
and FC curves and the hysteresis loops can be found not only in
regular ferromagnets \cite{Ashcroft} but also in spin glasses
\cite{Fischer}. Neutron scattering technique can resolve spatial
spin correlations and decisively distinguish the two cases. Our
neutron diffraction experiments on polycrystals
(Ba$_{0.7}$K$_{0.3}$)(Zn$_{0.9}$Mn$_{0.1}$)$_2$As$_2$ \cite{ZhaoK}
and (La$_{0.9}$Sr$_{0.1}$)(Zn$_{0.9}$Mn$_{0.1}$)AsO \cite{Lu} were
not able to decouple the magnetic and structural Bragg peaks since
they superpose to each other. Single crystals are expected for high
resolution neutron scattering experiments. None the less, the moment
size is usually small for typical spin glasses, i.e., $\sim$ 0.01
$\mu_B$/Mn for the II-VI (Zn,Mn)Se or other typical dilute alloy
spin glasses \cite{Tholence,Monod,Prejean}. Considering that the
saturation moment size of our
(Sr$_3$La$_2$O$_{5}$)(Zn$_{1-x}$Mn$_x$)$_2$As$_2$ is as large as 0.5
$\mu_B$, we tentatively assign it to a ferromagnetic ordering rather
than a spin glass.

The ferromagnetism in various diluted magnetic semiconductors and
oxides has been explained by several theoretical models, such as
Zener's model \cite{Dietl2}, percolation of bound magnetic polarons
(BMPs) \cite{Bhatt, Sarma, Coey}, and $d-d$ double exchange due to
hopping between transition metal $d$ states \cite{Millis}. In the
case of (Sr$_3$La$_2$O$_{5}$)(Zn$_{1-x}$Mn$_x$)$_2$As$_2$, the
resistivity is large ($\sim$ 10$^4$ $\Omega$ mm at 2 K) and the
carrier density is $\sim$ 10$^{16}$ cm$^{-3}$. The low carrier
density and the relative low $T_C$ seems more amenable with the BMPs
model.

\section{Summary}

In summary, we report the synthesis and characterization of bulk
form diluted magnetic semiconductor
(Sr$_3$La$_2$O$_{5}$)(Zn$_{1-x}$Mn$_x$)$_2$As$_2$ with the Curie
temperature $\sim$ 40 K. The physical properties of this layered
polycrystal are very similar to the recently discovered bulk form
DMSs, (La,Ba)(Zn,Mn)AsO \cite{Ding1} and
(Ba,K)(Zn,Mn)$_{2}$As$_{2}$\cite{ZhaoK}. Summing up previously
reported ``11" type (Zn,Mn)Se (II-VI), ``111" type Li(Zn,Mn)As,
``1111" type (La,Ba)(Zn,Mn)AsO and ``122" type
(Ba,K)(Zn,Mn)$_{2}$As$_{2}$,  the ``32522" type
(Sr$_3$La$_2$O$_{5}$)(Zn$_{1-x}$Mn$_x$)$_2$As$_2$ system in current
study represents the fifth DMS system that has a direct counterpart
with identical/similar structure in the Fe-based superconductors. As
we stated in earlier papers \cite{Deng1,Ding1, ZhaoK}, the common
structure and excellent lattice matching between ferromagnetic,
antiferrmagnetic and superconducting systems make it possible to
make junctions between these systems through the As layer.

\acknowledgments{The work at Zhejiang University was supported by
National Basic Research Program of China (No. 2011CBA00103, No.
2014CB921203), NSFC (No. 11274268); F.L. Ning acknowledges helpful
discussion with C.Q. Jin and Y.J. Uemura.}
\\


\begin{thebibliography}{00}


\bibitem{Ohno} \Name{Ohno H., Shen A., Matsukura F., Oiwa A., Endo A., Katsumoto S., and Iye Y.} \REVIEW{Appl. Phys. Lett.} {69}{1996}{363}.

\bibitem{Samarth} \Name{Samarth N.} \REVIEW{Nat. Mater.} {9}{2010}{955}.

\bibitem{Chambers} \Name{Chambers S.} \REVIEW{Nat. Mater.} {9}{2010}{956}.

\bibitem{Dietl1} \Name{Dietl T.} \REVIEW{Nat. Mater.} {9}{2010}{965}.

\bibitem{Zutic} \Name{Zutic I., Fabian J., and Das Sarma S.} \REVIEW{Rev. Mod. Phys.} {76}{2004}{323}.

\bibitem{Jungwirth} \Name{Jungwirth T., Sinova J., Masek J., Kucera J., and MacDonald A.H.} \REVIEW{Rev. Mod. Phys.} {78}{2006}{809}.

\bibitem{Deng1} \Name{Deng Z., Jin C.Q.,
  Liu Q.Q., Wang X.C., Zhu J.L., Feng S.M., Chen L.C., Yu R.C., Arguello C., Goko T., Ning F.L., Zhang J.S., Wang Y.Y., Aczel A.A., Munsie T., Williams T.J., Luke G.M., Kakeshita T., Uchida S., Higemoto W., Ito T.U., Gu B., Maekawa S., Morris G.D. and Uemura Y.J.} \REVIEW{Nat. Commun.} {2}{2010}{422}.

\bibitem{Deng2} \Name{Deng Z., Zhao K., Gu B., Han W., Zhu J.L., Wang X.C., Li X., Liu Q.Q., Yu R.C., Goko T., Frandsen B., Liu L., Zhang J.S., Wang Y.Y., Ning F.L., Maekawa S., Uemura Y.J. and Jin C.Q. } \REVIEW{Phys. Rev. B} {88}{2013}{081203(R)}.

\bibitem{Ding1} \Name{Ding C., Man H.Y., Qin C., Lu J.C., Sun Y.L., Wang Q., Yu B. Q., Feng C.M., Goko T., Arguello C.J., Liu L., Frandsen B.A., Uemura Y.J., Wang H.D., Luetkens H., Morenzoni E., Han W., Jin C.Q., Munsie T., Williams T.J., D¡¯Ortenzio R.M., Medina T., Luke G.M., Imai T.,  and Ning F.L.} \REVIEW{Phys. Rev. B} {88}{2013}{041102(R)}.

\bibitem{HanW} \Name{Han W., Zhao K., Wang X.C., Liu Q.Q., Ning F.L., Deng Z., Liu Y., Zhu J.L., Ding C., and Jin C.Q} \REVIEW{Science China-Physics, Mechanics,Astronomy} {56}{2013}{2026}.

\bibitem{Yangxj} \Name{Yang X.J., Li Y.K., Shen C.Y., Si B.Q., Sun Y.L., Tao Q., Cao G.H., Xu Z.A., and Zhang F.C.} \REVIEW{Appl. Phys. Lett.} {103}{2013}{022410}.

\bibitem{ZhaoK} \Name{Zhao K., Deng Z., Wang X.C., Han W. , Zhu J.L., Li X., Liu Q.Q., Yu R.C., Goko T., Frandsen B., Liu L., Ning F.L., Uemura Y.J., Dabkowska H., Luke G.M., Luetkens H., Morenzoni E., Dunsiger S.R., Senyshyn A., B\"{o}ni P., and Jin C.Q.} \REVIEW{Nat. Commun.} {4}{2013}{1442}.

\bibitem{Yang2} \Name{Yang X.J., Li Y.K., Zhang P., Luo Y.K., Chen Q., Feng C.M., Cao C., Dai J.H., Tao Q., Cao G.H., and Xu Z.A.} \REVIEW{J. Appl. Phys.} {114}{2013}{223905}.

\bibitem{ZhaoJH1} \Name{Chen L., Yan S., Xu P.F., Wang W.Z., Deng J.J., Qian X., Ji Y., and Zhao J.H.} \REVIEW{Appl. Phys. Lett.} {95}{2009}{182505}.

\bibitem{ZhaoJH2} \Name{Chen L., Yang X., Yang F.H., Zhao J.H., Misuraca J., Xiong P. and Molnar S.V.} \REVIEW{Nano Lett.} {11}{2011}{2584}.

\bibitem{Dunsiger} \Name{Dunsiger S.R., Carlo J.P., Goko T., Nieuwenhuys G., Prokscha T., Suter A., Morenzoni E., Chiba D., Nishitani Y., Tanikawa T., Matsukura F., Ohno H., Ohe J., MaekawaS. and Uemura Y.J.} \REVIEW{Nat. Mater.} {9}{2010}{299}.

\bibitem{Ding2} \Name{Ding C., Qin C., Man H.Y., Imai T. and Ning F.L.} \REVIEW{Phys. Rev. B} {88}{2013}{041108(R)}.

\bibitem{Kamihara} \Name{Kamihara Y., Watanabe T., Hirano M., and Hosono H.} \REVIEW{J. Am. Chem. Soc.} {130}{2008}{3296}.

\bibitem{Emery} \Name{Emery N., Wildman E.J., Skakle J.M.S., Mclaughlin A.C., Smith R.I. and Fitch A.N.} \REVIEW{Phy. Rev. B} {83}{2011}{144429}.

\bibitem{WuMK} \Name{Hsu F.C., Luo J.Y., Yeh K.W., Chen T.K., Huang T.W., Wu P.M., Lee Y.C., Huang Y.L., Chu Y.Y., Yan D.C., and Wu M.K.} \REVIEW{Proc. Natl. Acad. Sci. U.S.A.} {105}{2008}{14262}.

\bibitem{Twardowski} \Name{Twardowski A., Swagten H.J.M., and Jonge W. J. M. de} \REVIEW{Phys. Rev. B} {36}{1987}{7013}.

\bibitem{Ito} \Name{Ito T., Ito K., and Oka M.} \REVIEW{Japanese Journal of Applied Physics} {17}{1977}{371}.

\bibitem{Wangxc} \Name{Wang X.C., Liu Q.Q., Lv Y.X., Gao W.B., Yang L.X., Yu R.C., Li F.Y., Jin C.Q.} \REVIEW{Solid State Communications} {148}{2008}{538}.

\bibitem{Bronger} \Name{Bronger W., Mueller P., Hoeppner R., and Schuster H.U.} \REVIEW{Z. Anorg. Allg. Chem.} {539}{1986}{175}.

\bibitem{Rotter} \Name{Rotter M., Tegel M., and Johrendt D.} \REVIEW{Phys. Rev. Lett.} {101}{2008}{107006}.

\bibitem{Singh} \Name{Singh Y., Green M.A., Huang Q., Kreyssig A., McQueeney R.J., Johnston D.C., and Goldman A.I.} \REVIEW{Phys. Rev. B} {80}{2009}{100403(R)}.

\bibitem{Shirage} \Name{Shirage P.M., Kihou K., Lee C.H., Kito H., Eisaki H., and Iyo A.} \REVIEW{J. Am. Chem. Soc.} {133}{2011}{9630}.

\bibitem{Zhu} \Name{Zhu X.Y., Han F., Mu G., Cheng P., Shen B., Zeng B., and Wen H.H.} \REVIEW{Phys. Rev. B} {79}{2009}{220512(R)}.

\bibitem{WenHH} \Name{Zhu X.Y., Han F., Mu G., Zeng B., Cheng P., Shen B., and Wen H.H.} \REVIEW{Phys. Rev. B} {79}{2009}{024516}.

\bibitem{Ashcroft} \Name{Ashcroft N.W., and Mermin N.D.} \REVIEW{\textit{Solid State Physics}}  {}{Holt, Rinehart and Winston, 1976}{}.

\bibitem{Fischer} \Name{Fischer K.H.,and Hertz J.A.} \REVIEW{\textit{Spin Glasses}}  {}{Cambridge University Press, 1991}{}.

\bibitem{Lu} \Name{Lu J.C., Man H.Y., Ding C., Wang Q., Yu B.Q., Guo S.L., Uemura Y.J., Han W., Jin C.Q., Wang H.D., Chen B., and Ning F.L.} \REVIEW{Europhysics Letters} {103}{2013}{67011}.

\bibitem{Tholence} \Name{Tholence J.L., and Tournier R.} \REVIEW{J. Phys. (Paris)} {35}{1974}{C4-229}.

\bibitem{Monod} \Name{Monod P., Prejean J.J., and Tissier B.} \REVIEW{J. Appl. Physics} {50}{1979}{7324}.

\bibitem{Prejean} \Name{Prejean J.J., Joliclerc M. , and Monod P.} \REVIEW{J. Phys. (Paris)} {41}{1980}{427}.

\bibitem{Dietl2} \Name{Dietl T., Ohno H., Matsukura F., Cibert J., and Ferrand D.} \REVIEW{Science} {287}{2000}{1019}.

\bibitem{Bhatt} \Name{Berciu M., and Bhatt R.N.} \REVIEW{Phys. Rev. Lett.} {87}{2001}{107203}.

\bibitem{Sarma} \Name{Kaminski A. and Sarma S.D.} \REVIEW{Phys. Rev. Lett.} {88}{2002}{247202}.

\bibitem{Coey} \Name{Coey J.M.D., Venkatesan M. and Fitzgerald C.B.} \REVIEW{Nat. Mater.} {4}{2005}{173}.

\bibitem{Millis} \Name{Millis A.J.} \REVIEW{Nature} {392}{1998}{147}.




\end{thebibliography}
\end{document}